\documentclass[notitlepage,nofootinbib,preprintnumbers,aps,prd]{revtex4-1}
\usepackage{amsmath,amssymb,natbib,bm,color}
\usepackage{tikz}
\usetikzlibrary{positioning,decorations.pathmorphing,decorations.markings,shapes.symbols}
\usepackage{appendix}

\begin{document}

\title{The Effects of Primordial Black Holes on Dark Matter Models}
\author{Paolo Gondolo}
\author{Pearl Sandick}
\author{Barmak Shams Es Haghi}
\affiliation{Department of Physics and Astronomy, University of 
Utah, Salt Lake City, UT 84112, USA}
\begin{abstract}
We investigate the effects of producing dark matter by Hawking evaporation of primordial black holes (PBHs) in scenarios that may have a second well-motivated dark matter production mechanism, such as freeze-out, freeze-in, or gravitational production. We show that the interplay between PBHs and the alternative sources of dark matter can give rise to model-independent modifications to the required dark matter abundance from each production mechanism, which in turn affect the prospects for dark matter detection. In particular, we demonstrate that for the freeze-out mechanism, accounting for evaporation of PBHs after freeze-out demands a larger annihilation cross section of dark matter particles than its canonical value for a thermal dark matter. For mechanisms lacking thermalization due to a feeble coupling to the thermal bath, we show that the PBH contribution to the dark matter abundance leads to the requirement of an even feebler coupling. Moreover, we show that when a large initial abundance of PBHs causes an early matter-dominated epoch, PBH evaporation alone cannot explain the whole abundance of dark matter today.  In this case, an additional production mechanism is required, in contrast to the case when PBHs are formed and evaporate during a radiation-dominated epoch.
\end{abstract}

\maketitle

%%%%%%%%%%%%%%%%%%%%%%%%%%%%%%INTRODUCTION%%%%%%%%%%%%%%%%%%%%%%%%%%%%%%%%%%%%
%%%%%%%%%%%%%%%%%%%%%%%%%%%%%%%%%%%%%%%%%%%%%%%%%%%%%%%%%%%%%%%%%%%%%%%%%%%
%%%%%%%%%%%%%%%%%%%%%%%%%%%%%%%%%%%%%%%%%%%%%%%%%%%%%%%%%%%%%%%%%%%%%%%%%%%
\section{Introduction}

The nature and origin of dark matter still remains one of the main unresolved issues in particle physics, astrophysics, and cosmology. 
Despite considerable efforts to detect particle dark matter, all past searches, including direct detection experiments~\cite{Fu:2016ega, Aprile:2017iyp, Akerib:2016lao, Behnke:2016lsk, Akerib:2016vxi, Tan:2016zwf}, indirect detection experiments~\cite{Hooper:2010mq, Bulbul:2014sua, Urban:2014yda, Choi:2015ara, Ruchayskiy:2015onc, Ackermann:2015zua, Franse:2016dln, Aharonian:2016gzq, Cui:2016ppb, Aartsen:2016zhm, TheFermi-LAT:2017vmf}, and collider searches~\cite{Abercrombie:2015wmb, Aaboud:2016wna, Sirunyan:2018gka, Aaboud:2019yqu} have come short of observing a conclusive signal.

Each of these experiments targets dark matter particles within a specific mass range, motivated by particular dark matter production mechanisms such as freeze-out~\cite{Steigman:1984ac, Lee:1977ua, Hochberg:2014dra}, freeze-in~\cite{Hall:2009bx}, gravitational production~\cite{Chung:1998ua, Chung:1998zb, Chung:1998bt, Chung:2001cb, Chung:2004nh}, etc. The null results from current dark matter searches necessitate exploring novel dark matter production mechanisms. New mechanisms of dark matter production open up new directions in model building and, in turn, discovery may require novel dark matter experiments and techniques.

One intriguing dark matter production mechanism is the Hawking evaporation of a population of primordial black holes (PBHs).
The possibility of the formation of small black holes at the early stages of the universe and their implications in cosmology have been extensively studied since this idea was first introduced~\cite{Novikov:1967}.

The high density of the early universe provides a necessary condition for PBH formation, but it is not sufficient. Different well-studied scenarios such as collapse from inhomogeneities~\cite{Carr:1975qj,Nadezhin:1978}, sudden reduction in the pressure~\cite{Khlopov:1980mg, Khlopov:1981mg, Khlopov:1982mg}, collapse of cosmic loops~\cite{Hawking:1987bn, Polnarev:1988dh, Hansen:1999su, Hogan:1984zb, Nagasawa:2005hv}, bubble collisions~\cite{Crawford:1982yz, Hawking:1982ga, La:1989st, Moss:1994iq, Khlopov:1998nm, Konoplich:1999qq}, and collapse of domain walls~\cite{Rubin:2000dq, Rubin:2001yw, Dokuchaev:2004kr} can lead to PBH production in the early universe. For a recent review, see~\cite{Carr:2020gox}.

Small enough PBHs ($0.1\text{g}\lesssim M_\text{BH}\lesssim10^9\text{g}$) disappear before Big Bang Nucleosynthesis (BBN), and their abundance at formation time is not constrained. 
Due to the nontrivial behavior of quantum fields in the curved space-time background in the vicinity of a black hole, a black hole loses mass constantly by emitting all the particles that are lighter than its temperature. The resultant Hawking radiation of PBHs has been considered as a possible explanation for baryogenesis~\cite{Hawking:1974rv, Zeldovich:1976vw}, baryogenesis and dark matter~\cite{Baumann:2007yr, Fujita:2014hha, Morrison:2018xla}, and  dark matter~\cite{Lennon:2017tqq, Allahverdi:2017sks, Hooper:2019gtx, Masina:2020xhk, Baldes:2020nuv}. 

In this paper, we focus on the production of dark matter particles by PBHs when at least one other production mechanism is involved. The alternative mechanism can happen because of possible non-gravitational interactions of dark matter, e.g., freeze-out and freeze-in, or can be governed by purely gravitational interactions, e.g., freeze-in and gravitational production of superheavy dark matter. The combination of PBHs and another dark matter production mechanism leads to interesting model-independent modifications to relevant parameters of these production mechanisms that consequently affect the prospects for dark matter detection. 

Dark matter particles produced by PBHs might thermalize with the thermal bath or a preexisting population of thermal dark matter particles, provided that dark matter has strong enough non-gravitational interactions. Thermalization, which is only effective if PBHs disappear before the freeze-out of the dark matter, will erase the effect of PBHs on the abundance of the dark matter today. Evaporation of PBHs after freeze-out, on the other hand, contributes directly to the final abundance of the dark matter. To avoid overclosing the universe, freeze-out annihilation must be increased accordingly to produce less dark matter. The requisite larger annihilation cross section for dark matter implies a stronger signal in indirect searches for dark matter. 

On the other hand, very weakly interacting dark matter particles that come from PBHs cannot thermalize, and therefore contribute directly to the dark matter abundance today. In this case, the relevant production mechanisms, such as freeze-in or gravitational production of superheavy dark matter, must be reduced accordingly to produce less dark matter. For the freeze-in mechanism this means that the very weak interaction between dark matter and the thermal bath needs to become even weaker, and therefore the searches for these particles would become even more challenging.

The outline of this paper is as follows. In Section~\ref{sec:PBH} we review the formation and evaporation of PBHs in the early universe. Different mechanisms of dark matter production, including freeze-out, freeze-in, and gravitational production, and the possible interplay between them and PBH evaporation are discussed in Section~\ref{sec:DM}. 
In Section~\ref{sec:ps} we describe the available parameter space and relevant cosmological constraints.  Our results are presented in Section~\ref{sec:results}, and the general conclusions and outlook are discussed in Section~\ref{sec:conclusion}.

%%%%%%%%%%%%%%%%%%%%%PBHs FORMATION AND EVAPORATION%%%%%%%%%%%%%%%%%%%%%
%%%%%%%%%%%%%%%%%%%%%%%%%%%%%%%%%%%%%%%%%%%%%%%%%%%%%%%%%%%%%%%%%%%%%%%%%%%
%%%%%%%%%%%%%%%%%%%%%%%%%%%%%%%%%%%%%%%%%%%%%%%%%%%%%%%%%%%%%%%%%%%%%%%%%%%
\section{PBHs, formation and evaporation}
\label{sec:PBH}
In this section we review the formation and evaporation of PBHs.

In the early universe, density fluctuations $\delta\rho/\rho$ grow after they enter the cosmological horizon. If $\delta\rho/\rho$ is greater than the equation of state parameter $w\equiv p/\rho$, the fluctuation can collapse into a PBH with mass bounded by the total mass within the horizon~\cite{Carr:1975qj}. To overcome the pressure, the size of the overdense region must be larger than the Jeans length, which is $\sqrt{w}$ times the horizon size~\cite{Carr:1975qj, Carr:1974nx} (we assume a radiation-dominated epoch at the time of PBH formation, thus $w>0$).
Therefore the mass of a PBH formed in the radiation-dominated epoch is evaluated as
\begin{equation}
M_i=\frac{4\pi}{3}\gamma \rho_\text{rad}(T_i)  H(T_i)^{-3},
\label{eq:initialmass}
\end{equation}
where $\gamma\sim w^{3/2}\approx0.2$ in a radiation-dominated universe, $T_i$ is the temperature of the universe at black hole formation time, $H(T)=\sqrt{4\pi^3g_*(T)/45}\,T^2/M_\text{Pl}$ is the Hubble expansion rate, and $\rho_\text{rad}(T)$ is the energy density of the universe, given by
\begin{equation}
\rho_\text{rad}(T)=\frac{\pi^2}{30}g_*(T)T^4,~~~~~~g_*(T)=\sum_B g_B\left(\frac{T_B}{T}\right)^4+\frac{7}{8}\sum_F g_F\left(\frac{T_F}{T}\right)^4.
\end{equation} 
Here $g_*(T)$ denotes the total number of relativistic degrees of freedom, $T$ is the temperature of the universe, and the sum includes all the bosonic ($B$) and fermionic ($F$) degrees of freedom with temperatures respectively equal to $T_B$ and $T_F$.
Eq.~(\ref{eq:initialmass}) relates the initial mass of PBHs to their time of formation.

A black hole loses its mass by emitting particles that are lighter than its temperature via Hawking radiation~\cite{Hawking:1974sw}. Ignoring greybody factors, the Hawking radiation can be described as black body radiation. Thus the energy spectrum of the $i$th emitted species by a non-rotating black hole with zero charge is given by
\begin{equation}
\frac{d^2u_i(E,t)}{dtdE}=\frac{g_i}{8\pi^2}\frac{E^3}{e^{E/T_\text{BH}}\pm1} ,
\label{eq:rate}
\end{equation}
($+$ for fermion emission and $-$ for boson emission) where $u_i(E,t)$ is the total radiated energy per unit area, $g_i$ counts the number of degrees of freedom of the $i$th species, $E$ is the energy of the emitted particle, and $T_\text{BH}$ is the horizon temperature of the black hole, 
\begin{equation}
T_\text{BH}=\frac{M_\text{Pl}^2}{8\pi M_\text{BH}}.
\label{eq:temp}
\end{equation}
Eqs.~(\ref{eq:rate}) and (\ref{eq:temp}) can be used to find the mass loss rate of a black hole due to Hawking evaporation, which is
\begin{equation}
\frac{dM_\text{BH}}{dt}=-4\pi r_\text{S}^2\sum_i\int_0^\infty  \frac{d^2u_i(E,t)}{dtdE} dE=-\frac{g_*(T_\text{BH})}{30720 \pi}\frac{M_\text{Pl}^4}{M_\text{BH}^2} .
\label{eq:massloss}
\end{equation}
Here $g_*(T_\text{BH})$ denotes the total number of relativistic degrees of freedom emitted by the black hole, and $ r_\text{S}=2 M_\text{BH}/M_\text{Pl}^2$  is the Schwarzschild radius 
of the black hole.
Integrating Eq.~(\ref{eq:massloss}) gives the time evolution of the mass of a black hole with initial mass $M_i$ formed at  $t_i$,
\begin{equation}
M(t)=M_i\left(1-\frac{t-t_i}{\tau}\right)^{1/3},
\end{equation}
where
\begin{equation}
\tau=\frac{10240\pi}{g_*(T_\text{BH})}\frac{M_i^3}{M_\text{Pl}^4}
\end{equation}
is the black hole lifetime.

We calculate the temperature of the universe at the time of evaporation of PBHs, defined as 
$T_\text{eva}\equiv T(t_i+\tau)$. We first show that $t_i\ll\tau$, and thus $T_\text{eva} \simeq T(\tau)$. In the radiation-dominated epoch $H(t)=1/(2t)$. Combining this with the Friedmann equation, $H^2=8\pi\rho/3M_\text{Pl}^2$, and using Eq.~(\ref{eq:initialmass}), we can easily see that 
\begin{equation}
\frac{t_i}{\tau}=\frac{g_*(T_\text{BH})}{10240\pi\gamma}\left(\frac{M_\text{Pl}}{M_\text{BH}}\right)^2\simeq7.6\times10^{-12}\left(\frac{g_*(T_\text{BH})}{106.8}\right)\left(\frac{0.2}{\gamma}\right)\left(\frac{1 \text{g}}{M_\text{BH}}\right)^2 .
\end{equation}
Therefore 
\begin{equation}
T_\text{eva} =  T(t_i+\tau)\simeq T(\tau)=\frac{\sqrt{3}\,g_*(T_\text{BH})^{1/4}}{64\sqrt{2}\,5^{1/4}\pi^{5/4}}\frac{M_\text{Pl}^{5/2}}{M_\text{BH}^{3/2}} .
\label{eq:Teva}
\end{equation}

Given the energy spectrum of the $i$th emitted species, its rate of emission per energy interval can be expressed as
\begin{equation}
\frac{d^2N_i}{dtdE}=\frac{4\pi r_\text{S}^2}{E}\frac{d^2u_i}{dtdE}=\frac{g_i}{2\pi}\frac{r_\text{S}^2E^2}{e^{E/T_\text{BH}}\pm1} .
\label{eq:numrate}
\end{equation}
By integrating Eq.~(\ref{eq:numrate}) over energy and time we calculate the total number of particles of the $i$th  species emitted over the lifetime of the black hole. For bosons,
\begin{eqnarray}
\nonumber N_i&=&\frac{120\,\zeta(3)}{\pi^3}\frac{g_i}{g_*(T_\text{BH})}\frac{M_\text{BH}^2}{M_\text{Pl}^2},~~~~~~T_\text{BH}>m_i , \\
N_i&=&\frac{15\,\zeta(3)}{8\pi^5}\frac{g_i}{g_*(T_\text{BH})}\frac{M_\text{Pl}^2}{m_i^2},~~~~~~~T_\text{BH}<m_i.
\label{eq:number}
\end{eqnarray}
The total number of fermionic species is $N_F=\frac{3}{4}\frac{g_F}{g_B}N_B$.

%%%%%%%%%%%%%%%%%%%%%%DARK MATTER PRODUCTION%%%%%%%%%%%%%%%%%%%%
%%%%%%%%%%%%%%%%%%%%%%%%%%%%%%%%%%%%%%%%%%%%%%%%%%%%%%%%%%%%%%%%%%%%%%%%%%%
%%%%%%%%%%%%%%%%%%%%%%%%%%%%%%%%%%%%%%%%%%%%%%%%%%%%%%%%%%%%%%%%%%%%%%%%%%%
\section{Dark Matter production}
\label{sec:DM}
Many mechanisms of dark matter production have been proposed to explain the observed dark matter abundance, to open up new avenues in model building, and to provide new directions in dark matter searches. Whatever the true nature of dark matter is, a population of PBHs may also contribute to the abundance of dark matter today due to Hawking evaporation. In this section we first discuss dark matter production by PBHs, then we review some other highly motivated dark matter production mechanisms, including freeze-out, freeze-in, and gravitational production of very heavy dark matter particles (e.g., WIMPZILLAs). We also discuss the possible interplay between dark matter production by PBHs and these other mechanisms.

%%%%%%%%%%%%%%%%%%%%%%DARK MATTER PRODUCTION BY PBHs%%%%%%%%%%%%%%%%%%%%
%%%%%%%%%%%%%%%%%%%%%%%%%%%%%%%%%%%%%%%%%%%%%%%%%%%%%%%%%%%%%%%%%%%%%%%%%%%
%%%%%%%%%%%%%%%%%%%%%%%%%%%%%%%%%%%%%%%%%%%%%%%%%%%%%%%%%%%%%%%%%%%%%%%%%%%
\subsection{Dark Matter Production by PBHs}
\label{subsec:DMfromPBH}
As we discussed in Section~\ref{sec:PBH}, a PBH loses mass through Hawking evaporation and emits particles that have mass smaller than its temperature. Since it is a purely gravitational process, 
Hawking radiation consists of all different species of particles, including dark matter.
The abundance today can be related to the total number of produced particles and the temperature of the universe at the time of the formation of PBHs via conservation of entropy. The amount of dark matter produced by Hawking evaporation of PBHs  in a radiation-dominated era is therefore
\begin{equation}
Y_\chi=\frac{n_\chi(T_0)}{s(T_0)}=\frac{n_\chi(T_\text{eva})}{s(T_\text{eva})}=N_\chi\frac{n_\text{BH}(T_i)}{s(T_i)},
\label{eq:yent}
\end{equation}
where a subscript $0$ means the quantity is evaluated today, $n_\chi(T)$ and $n_\text{BH}(T)$ are the number densities of dark matter particles $\chi$ and PBHs at radiation temperature $T$, respectively, $N_\chi$ is the total number of particles $\chi$ emitted during the lifetime of the PBH (Eq.~(\ref{eq:number})), and $s(T)$ is the entropy density given by
\begin{equation}
s(T)=\frac{2\pi^2}{45}g_{*,s}(T)T^3,~~~~~~g_{*,s}(T)=\sum_B g_B\left(\frac{T_B}{T}\right)^3+\frac{7}{8}\sum_F g_F\left(\frac{T_F}{T}\right)^3.
\label{eq:entropy}
\end{equation}

It is customary to introduce the dimensionless parameter $\beta$ to represent the initial energy density of PBHs normalized to the radiation energy density at the time of formation,
\begin{equation}
\beta=M_\text{BH}\frac{n_\text{BH}(T_i)}{\rho_\text{rad}(T_i)}.
\label{eq:beta}
\end{equation}
In this study $\beta$ is a free parameter.
By using the definition of $\beta$ and the fact that at high temperatures $g_{*,s}(T)\simeq g_*(T)$, we have
\begin{equation}
Y_\chi=\beta N_\chi\frac{\rho_\text{rad}(T_i)}{s(T_i)M_\text{BH}}=\frac{3\beta N_\chi}{4}\frac{T_i}{M_\text{BH}}.
\end{equation}
From Eq.~(\ref{eq:initialmass}), $T_i$ can be expressed in terms of the black hole initial mass, leading to
\begin{equation}
Y_\chi=\frac{3\sqrt{3}\,5^{1/4}}{8\pi^{3/4}}\beta N_\chi\gamma^{1/2}g_*(T_i)^{-1/4}\left(\frac{M_\text{pl}}{M_\text{BH}}\right)^{3/2}.
\label{eq:yield}
\end{equation}
The relic abundance of dark matter can be obtained as
\begin{equation}
\Omega_\chi=\frac{\rho_{\chi,0}}{\rho_c}=\frac{m_\chi Y_\chi }{\rho_c}s_0,
\label{eq:relicdef}
\end{equation}
with $\rho_c$ equal to the critical energy density of the universe.
By combining Eqs.~(\ref{eq:relicdef}),  (\ref{eq:yield}) and (\ref{eq:number}), we find the relic abundance of a bosonic dark matter particle when PBHs evaporate during a radiation-domianted era to be
\begin{eqnarray}
\nonumber\Omega_\chi&=&\frac{45\sqrt{3}\,5^{1/4}\zeta(3)}{\pi^{15/4}}\left(\frac{g_\chi}{g_*(T_\text{BH})}\right)g_*(T_i)^{-1/4}\beta \gamma^{1/2}\frac{m_\chi s(T_0)}{\rho_c}\left(\frac{M_\text{BH}}{M_\text{Pl}}\right)^{1/2}, ~~~~~~T_\text{BH}>m_\chi,\\
\Omega_\chi&=&\frac{45\sqrt{3}\,5^{1/4}\zeta(3)}{64\pi^{23/4}}\left(\frac{g_\chi}{g_*(T_\text{BH})}\right)g_*(T_i)^{-1/4}\beta \gamma^{1/2}\frac{m_\chi s(T_0)}{\rho_c}\left(\frac{M_\text{Pl}^7}{M_\text{BH}^3m_\chi^4}\right)^{1/2}, ~~~~~~T_\text{BH}<m_\chi.
\label{eq:relic1}
\end{eqnarray}
By inserting numerical values for $\rho_c=1.0537\times 10^{-5}\, h^2\,\,\rm{GeV}~{\rm cm}^{-3}$ and $s_0=2891.2\left(\frac{T_0}{2.7255}\right)^3 \rm{cm^{-3}}$~\cite{Aghanim:2018eyx} into Eq.~(\ref{eq:relic1}) the abundance of bosonic dark matter particles can be expressed as
\begin{eqnarray}
\nonumber\Omega_\chi h^2&\simeq &7.3 \times 10^7\beta\left(\frac{g_*(T_i)}{106.8}\right)^{-1/4}\left(\frac{\gamma}{0.2}\right)^{1/2}\left(\frac{m_\chi}{\rm{GeV}}\right)\left(\frac{g_\chi}{g_*(T_\text{BH})}\right)\left(\frac{M_\text{BH}}{M_\text{Pl}}\right)^{1/2}, ~~~~~~T_\text{BH}>m_\chi,\\
\Omega_\chi h^2&\simeq &1.2 \times 10^5\beta\left(\frac{g_*(T_i)}{106.8}\right)^{-1/4}\left(\frac{\gamma}{0.2}\right)^{1/2}\left(\frac{m_\chi}{\rm{GeV}}\right)\left(\frac{g_\chi}{g_*(T_\text{BH})}\right)\left(\frac{M_\text{Pl}^7}{M_\text{BH}^3m_\chi^4}\right)^{1/2}, ~~~~~~T_\text{BH}<m_\chi.
\label{eq:relic}
\end{eqnarray}
The relic abundance of a fermionic dark matter particle $F$ is related to that of a bosonic particle $B$ through $\Omega_F=\frac{3}{4}\frac{g_F}{g_B}\Omega_B$. These equations apply to dark matter produced by PBH evaporation during a radiation-dominated era.

The presence of PBHs may lead to an early matter-dominated era, in which case Eqs.~(\ref{eq:relic1}) and~(\ref{eq:relic}) need to be modified as follows. Since $\rho_\text{PBH}\propto a^{-3}$ and $\rho_\text{rad}\propto a^{-4}$, where $a$ is the scale factor, $\rho_\text{PBH}(t)/\rho_\text{rad}(t)$ grows with the expansion of the universe. Therefore, an initially radiation-dominated universe will eventually become matter-dominated if the PBHs are still around. The critical initial abundance of PBHs $\beta_c$ that leads to an early matter-dominated era can be obtained by demanding that PBH evaporation happens after an early equality time $t_\text{early-eq}$, defined by $\rho_\text{PBH}(t_\text{early-eq})/\rho_\text{rad}(t_\text{early-eq})\sim 1$.
This early equality time (or equivalent temperature) can be expressed in terms of $T_i$ and $\beta_c$,
\begin{equation}
\frac{\rho_\text{PBH}(T_\text{early-eq})}{\rho_\text{rad}(T_\text{early-eq})}=\frac{\rho_\text{PBH}(T_i)}{\rho_\text{rad}(T_i)}\frac{T_i}{T_\text{early-eq}}=\beta_c\frac{T_i}{T_\text{early-eq}}\sim 1.
\end{equation}
An early matter-dominated era is inevitable if $t_\text{early-eq}\lesssim t_\text{eva}$, or equivalently when
\begin{equation}
\beta>\beta_c=\frac{T_\text{eva}}{T_i}=\sqrt{\frac{g_*(T_\text{BH})}{10240\pi\gamma}}\frac{M_\text{Pl}}{M_\text{BH}}\simeq2.8\times 10^{-6}\left(\frac{g_*(T_\text{BH})}{106.8}\right)^{1/2}\left(\frac{0.2}{\gamma}\right)^{1/2}\left(\frac{1 \text{g}}{M_\text{BH}}\right) .
\label{eq:MD}
\end{equation}
Here $T_i$ and $T_\text{eva}$ are calculated from Eqs.~(\ref{eq:initialmass}) and~(\ref{eq:Teva}).

The expression for the relic abundance of dark matter given by Eq.~(\ref{eq:relic}) is valid as long as PBH evaporation happens in a radiation-dominated era. In an early matter-dominated era caused by PBHs, the entropy of the universe at the time of evaporation is determined by the PBH evaporation products. Because of this entropy production, the abundance of dark matter particles coming from Hawking evaporation of PBHs turns out to be independent of $\beta$~\cite{Baumann:2007yr}. The amount of dark matter produced by PBHs during such a matter-dominated era is
\begin{equation}
Y_\chi=\frac{n_\chi(T_0)}{s(T_0)}=\frac{n_\chi(T_\text{RH-BH})}{s(T_\text{RH-BH})}=N_\chi\frac{n_\text{BH}(\tau)}{s(T_\text{RH-BH})},
\end{equation}
where $T_\text{RH-BH}$ denotes the temperature that the radiated particles from PBHs equilibrate to, assuming an instantaneous thermalization. 
Since in the matter-dominated epoch $H(t)=2/(3t)$, the Friedmann equation $H^2=8\pi\rho_\text{BH}/(3M_\text{Pl}^2)$ can be used to obtain $\rho_\text{BH}(\tau)$ or equivalently $n_\text{BH}(\tau)$ in terms of $M_\text{BH}$, which gives
\begin{equation}
Y_\chi=\frac{3 g_*(T_\text{RH-BH})^{1/4}}{128\sqrt{2}\,5^{1/4}\pi^{5/4}}N_\chi\left(\frac{M_\text{Pl}}{M_\text{BH}}\right)^{5/2},
\label{eq:yieldMD}
\end{equation}
assuming that at high temperatures $g_{*,s}(T)\simeq g_*(T)$. 

Using Eqs.~(\ref{eq:relicdef}),  (\ref{eq:yieldMD}) and (\ref{eq:number}), the relic abundance of a bosonic dark matter particle when PBHs dominate the energy density of the universe (early matter-dominated era) is therefore
\begin{eqnarray}
\nonumber\Omega_\chi&=&\frac{9\times5^{3/4}\zeta(3)}{16\sqrt{2}\,\pi^{17/4}}\left(\frac{g_\chi}{g_*(T_\text{BH})}\right)g_*(T_\text{RH-BH})^{1/4}\frac{m_\chi s(T_0)}{\rho_c}\left(\frac{M_\text{Pl}}{M_\text{BH}}\right)^{1/2}, ~~~~~~T_\text{BH}>m_\chi,\\
\Omega_\chi&=&\frac{9\times5^{3/4}\zeta(3)}{1024\sqrt{2}\,\pi^{25/4}}\left(\frac{g_\chi}{g_*(T_\text{BH})}\right)g_*(T_\text{RH-BH})^{1/4}\frac{m_\chi s(T_0)}{\rho_c}\left(\frac{M_\text{Pl}^9}{M_\text{BH}^5m_\chi^4}\right)^{1/2}, ~~~~~~T_\text{BH}<m_\chi.
\label{eq:relicMD}
\end{eqnarray}
After inserting numerical values for $\rho_c$ and $s_0$ into Eq.~(\ref{eq:relicMD}), we have
\begin{eqnarray}
\nonumber\Omega_\chi h^2&\simeq & 1.1\times10^{7}\left(\frac{g_*(T_\text{RH-BH)}}{106.8}\right)^{1/4}\left(\frac{g_\chi}{g_*(T_\text{BH})}\right)\left(\frac{m_\chi}{\rm{GeV}}\right)\left(\frac{M_\text{Pl}}{M_\text{BH}}\right)^{1/2}, ~~~~~~T_\text{BH}>m_\chi,\\
\Omega_\chi h^2&\simeq & 1.7\times10^{4}\left(\frac{g_*(T_\text{RH-BH})}{106.8}\right)^{1/4}\left(\frac{g_\chi}{g_*(T_\text{BH})}\right)\left(\frac{m_\chi}{\rm{GeV}}\right)\left(\frac{M_\text{Pl}^9}{M_\text{BH}^5m_\chi^4}\right)^{1/2}, ~~~~~~T_\text{BH}<m_\chi.
\label{eq:relicMD2}
\end{eqnarray}
As usual, the relic abundance of a fermionic dark matter particle $F$ is related to that of a bosonic particle $B$ through $\Omega_F=\frac{3}{4}\frac{g_F}{g_B}\Omega_B$.

%%%%%%%%%%%%%%%%%%%%%%%%%%%FREEZE-OUT%%%%%%%%%%%%%%%%%%%%%%%%%%%%%%%%%%%%%
%%%%%%%%%%%%%%%%%%%%%%%%%%%%%%%%%%%%%%%%%%%%%%%%%%%%%%%%%%%%%%%%%%%%%%%%
%%%%%%%%%%%%%%%%%%%%%%%%%%%%%%%%%%%%%%%%%%%%%%%%%%%%%%%%%%%%%%%%%%%%%%%
\subsection{Freeze-out Production}
The most popular and motivated dark matter production mechanism is thermal freeze-out. Although freeze-out is usually discussed in the context of weakly interacting massive particles (WIMPs), in this study we use it in a broader sense to also include strongly interacting massive particles (SIMPs). More precisely, by freeze-out we mean reaching a constant comoving dark matter density by dropping out of chemical equilibrium either between dark matter particles and the thermal bath (WIMP case) or among dark matter particles themselves (SIMP case).

%%%%%%%%%%%%%%%%%%%%%%%%%%%WIMP%%%%%%%%%%%%%%%%%%%%%%%%%%%%%%%%%%%%%
%%%%%%%%%%%%%%%%%%%%%%%%%%%%%%%%%%%%%%%%%%%%%%%%%%%%%%%%%%%%%%%%%%%%%%%%
%%%%%%%%%%%%%%%%%%%%%%%%%%%%%%%%%%%%%%%%%%%%%%%%%%%%%%%%%%%%%%%%%%%%%%%
\subsubsection{WIMP}
WIMPs can reach thermal equilibrium with the thermal bath in the early universe through annihilation and pair production processes. As the temperature of the bath drops below the mass of the dark matter particles, the equilibrium dark matter abundance is suppressed exponentially by the temperature to mass ratio until the rate of  production becomes slower than the expansion rate of the universe; after that the comoving number density of dark matter particles remains constant. 
In this scenario dark matter attains thermal equilibrium through interaction ($\chi\chi\leftrightarrow BB$) where $B$ is a bath particle.
The number density of dark matter evolves according to the Boltzmann equation
\begin{equation}
\dot{n}_\chi+3Hn_\chi=-\langle\sigma v\rangle\left(n_\chi^2-n_{\chi,\text{eq}}^2\right),
\end{equation}
where  $\langle\sigma v\rangle$ is the thermally averaged annihilation cross section of the WIMPs.
It can be shown that~\cite{Kolb:1990vq} freeze-out happens when $x_f\equiv m_\chi/T\sim20$ (with a logarithmic dependence on $m_\chi$ and $\langle\sigma v\rangle$) and the WIMP abundance today for an $s$-wave annihilation cross section is given by
\begin{equation}
\Omega_\chi^{\text{fr-out}}(\langle\sigma v\rangle)h^2\simeq\frac{3.79\,x_f}{g_*^{1/2}}\frac{s_0}{M_\text{Pl}\rho_c\langle\sigma v\rangle}h^2. 
\end{equation}
We note that more complex dark matter models may include additional relevant effects, such as coannihilation with other bath particles, but the main idea remains unchanged~\cite{Griest:1990kh}.

Interestingly, a typical weak scale cross section leads to the right relic abundance. Although this approximate scheme is valid in principle for any dark matter mass, partial wave unitarity provides an upper bound on the annihilation cross section or equivalently on the WIMP mass~\cite{Griest:1989wd}. In the low-velocity limit where the cross section is assumed to be $s$-wave dominated, $\langle\sigma v\rangle\leq4\pi/(m_\chi^2 v)$, with $v\sim(6/x_f)^{1/2}$ for annihilation at freeze-out.
The observed abundance of cold dark matter, $\Omega_\text{c}h^2=0.12$~\cite{Aghanim:2018eyx}, leads to the upper bound $m_\chi \lesssim10^5$ GeV.

The possibility of formation of PBHs in the early universe and their subsequent Hawking evaporation into dark matter particles provides a second source of WIMP dark matter in addition to thermal production. If the evaporation of PBHs ends before WIMP freeze-out, then the dark matter particles produced by Hawking evaporation can reach chemical equilibrium with the bath (due to the enormous number of dark matter particles produced thermally), and the WIMPs from PBH evaporation give no extra contribution to the final relic abundance of the dark matter. On the other hand, if PBHs evaporate after the freeze-out of the WIMP, then the dark matter particles produced by Hawking evaporation can neither thermalize with the bath nor annihilate efficiently with each other, and will therefore directly contribute to the final dark matter density. To validate this statement, we compare the rate of annihilation of dark matter particles produced by Hawking evaporation with the Hubble rate at the evaporation time. The number density of dark matter particles produced by Hawking evaporation can be evaluated as $n_\chi(T_\text{eva})=s(T_\text{eva})Y_\chi$. By using Eqs.~(\ref{eq:yield}), (\ref{eq:entropy}), (\ref{eq:number}), (\ref{eq:Teva}), and estimating the rate of annihilation of Majorana fermion dark matter particles by $\Gamma\sim n_\chi(T_\text{eva})\langle\sigma v\rangle$ with $\langle\sigma v\rangle\sim1/\bar E^2$ (where  $\bar E\simeq 5.4\,T_\text{BH}$ when $T_\text{BH}>m_\chi$ and $\bar E\simeq 5.4\,m_\chi$ when $T_\text{BH}<m_\chi$) we have
\begin{eqnarray}
\nonumber\Gamma/H(T_\text{eva})&\sim&\frac{27\sqrt{5}\zeta(3)\beta\sqrt{\gamma}}{4\sqrt{2}\pi^{5/2}}\frac{g_\chi}{g_*(T_\text{eva})^{1/2}}\frac{M_\text{BH}}{M_\text{Pl}}\\
\nonumber&\simeq&3\times10^{3}\beta\left(\frac{\gamma}{0.2}\right)^{1/2}\left(\frac{106.8}{g_*(T_\text{eva})}\right)^{1/2}\left(\frac{g_\chi}{2}\right)\left(\frac{M_\text{BH}}{1 \text{g}}\right),~~~~~~(T_\text{BH}>m_\chi),\\
\nonumber\Gamma/H(T_\text{eva})&\sim&\frac{27\sqrt{5}\zeta(3)\beta\sqrt{\gamma}}{16384\sqrt{2}\pi^{13/2}}\frac{g_\chi}{g_*(T_\text{eva})^{1/2}}\frac{M_\text{Pl}^7}{M_\text{BH}^3m_\chi^4}\\
&\simeq&3.6\times10^{55}\beta\left(\frac{\gamma}{0.2}\right)^{1/2}\left(\frac{106.8}{g_*(T_\text{eva})}\right)^{1/2}\left(\frac{g_\chi}{2}\right)\left(\frac{1 \text{g}}{M_\text{BH}}\right)^3\left(\frac{1 \text{GeV}}{m_\chi}\right)^4,~~~~~~(T_\text{BH}<m_\chi),
\end{eqnarray}
where $g_{*,s}(T)\simeq g_*(T)$ at high temperatures is used. As we will see later (Fig.~\ref{fig:fig2}), even when the whole abundance of dark matter today is explained by Hawking evaporation of PBHs, the corresponding $\beta$ is too small to make the annihilation rate noticeable. Thus dark matter particles produced by PBHs cannot annihilate efficiently and this effect can be safely ignored.

Therefore, the effect of PBH evaporation on the WIMP abundance can be summarized as follows:
\begin{eqnarray}
\nonumber\Omega_\chi^{\text{fr-out}}(\langle\sigma v\rangle)h^2&\leq&\Omega_\text{c}h^2,~~~~~~~(t_{\text{eva}}<t_{\text{fr-out}}),\\
\Omega_\chi^{\text{fr-out}}(\langle\sigma v\rangle)h^2&+&\Omega_\chi^{\text{BH}}(m_\chi,M_\text{BH},\beta)h^2\leq\Omega_\text{c}h^2,~~~~~~~(t_{\text{eva}}>t_{\text{fr-out}}).
\end{eqnarray}
Dark matter particles originating from PBH evaporation after WIMP freeze-out make it possible to reduce the contribution of thermally produced WIMPs, which is equivalent to allowing the annihilation cross section to be larger. In fact, the annihilation cross section can easily saturate the unitarity bound. 

Although a larger annihilation cross section 
would be more easily detectable by indirect dark matter searches, there are robust limits on WIMP annihilation from $\it{Planck}$ measurements of the Cosmic Microwave Background (CMB)~\cite{Aghanim:2018eyx}, $\it{Fermi}$ measurements of Dwarf Spheroidal Galaxies of the Milky Way~\cite{Ackermann:2015zua, Fermi-LAT:2016uux}, the Alpha Magnetic Spectrometer (AMS) measurements of cosmic-rays~\cite{Aguilar:2014mma, Accardo:2014lma}, and neutrino experiments such as IceCube~\cite{Aartsen:2016pfc, Aartsen:2017ulx}. Ref.~\cite{Leane:2018kjk} provides model-independent limits on the WIMP annihilation cross section for s-wave $2\rightarrow 2$ annihilation to visible final states. Updated constraints on dark matter annihilation into neutrinos can be found, for example, in~\cite{Arguelles:2019ouk}.

Searches for dark matter at direct detection experiments and colliders also set strong limits on WIMP properties, but because of the model-dependent nature of these searches, we focus on indirect detection experiments.

%%%%%%%%%%%%%%%%%%%%%%%%%%%SIMP%%%%%%%%%%%%%%%%%%%%%%%%%%%%%%%%%%%%%
%%%%%%%%%%%%%%%%%%%%%%%%%%%%%%%%%%%%%%%%%%%%%%%%%%%%%%%%%%%%%%%%%%%%%%%%
%%%%%%%%%%%%%%%%%%%%%%%%%%%%%%%%%%%%%%%%%%%%%%%%%%%%%%%%%%%%%%%%%%%%%%%
\subsubsection{SIMP}
Generally SIMP couplings to the thermal bath are so weak that the dominant number-changing process for them is the $3\rightarrow 2$ self-annihilation rather than the $2\rightarrow 2$ annihilation, but they are large enough to keep SIMPs in kinetic equilibrium with the thermal bath. Therefore the thermal production mechanism for SIMPs is based on the freeze-out via $3\rightarrow 2$ self-annihilation of the dark matter~\cite{Hochberg:2014dra}.
The number density of dark matter in this scenario changes according to the Boltzmann equation,
\begin{equation}
\dot{n}_\chi+3Hn_\chi=-\langle\sigma v^2\rangle\left(n_\chi^3-n_\chi^2n_{\chi,\text{eq}}\right),
\end{equation}
where  $\langle\sigma v^2\rangle$ parameterizes the $3\rightarrow 2$ annihilation cross section of the SIMP.
It can be shown that freeze-out happens when $x_f\equiv m_\chi/T\sim20$ and the SIMP abundance today for a constant annihilation cross section is given by~\cite{Choi:2017mkk}
\begin{equation}
\Omega_\chi^{\text{fr-out}}(m_\chi,\langle\sigma v^2\rangle)h^2\simeq\frac{3\sqrt{3}\,5^{3/4} x_f^2}{\pi^{5/4}g_*^{3/4}}\frac{s_0}{\sqrt{M_\text{Pl}}m_\chi\rho_c\sqrt{\langle\sigma v^2\rangle}}h^2.
\end{equation}
In analogy to the WIMP case, where weak-scale couplings point to the weak scale, SIMP freeze-out points to strong-scale dark matter with strong couplings~\cite{Hochberg:2014dra}. 

Similar to the WIMP case, by adding PBH evaporation to this picture, there are two sources of WIMP production, the thermal production
and PBH evaporation. If evaporation of PBHs ends before SIMP freeze-out, then the dark matter particles produced by Hawking evaporation thermalize with the dark matter particles that are in thermal equilibrium with the bath (due to the large number of thermally produced dark matter particles) and they have no effect on the final relic abundance of SIMP dark matter. On the other hand, if PBHs evaporate after SIMP freeze-out, then the dark matter particles produced by Hawking evaporation can neither thermalize among themselves nor with the thermally produced dark matter particles, therefore they will be added to the relics of the freeze-out process. 
Hence, the effect of PBH evaporation on the SIMP abundance can be summarized as follows:
\begin{eqnarray}
\nonumber\Omega_\chi^{\text{fr-out}}(m_\chi, \langle\sigma v^2\rangle)h^2&\leq&\Omega_\text{c}h^2,~~~~~~~(t_{\text{eva}}<t_{\text{fr-out}}),\\
\Omega_\chi^{\text{fr-out}}(m_\chi, \langle\sigma v^2\rangle)h^2&+&\Omega_\chi^{\text{BH}}(m_\chi,M_\text{BH},\beta)h^2\leq\Omega_\text{c}h^2,~~~~~~~(t_{\text{eva}}>t_{\text{fr-out}}).
\end{eqnarray}

As in the WIMP case, SIMPs generated by PBH evaporation after freeze-out necessitate a reduction in the abundance of thermally produced dark matter particles, or equivalently, a larger self-annihilation cross section. 
Dark matter self-interactions have been motivated by the potential to resolve tensions between small-scale structure observations and N-body simulations of collisionless cold dark matter. 
Problems such as the ``cusp vs core problem''~\cite{Flores:1994gz, Moore:1994yx, Oh:2010mc} and the ``too-big-to-fail problem''~\cite{BoylanKolchin:2011de, BoylanKolchin:2011dk, Garrison-Kimmel:2014vqa, Papastergis:2014aba} can possibly be explained by a sizable self-interaction among dark matter particles~\cite{BoylanKolchin:2011de, Spergel:1999mh, deBlok:2009sp}. More precisely, dark matter self-scattering cross sections, $\sigma_\text{scatter}$, in the range $\sigma_\text{scatter}/m_\chi \gtrsim(0.1-2)\, \text{cm}^2/\text{g}$~\cite{Vogelsberger:2012ku, Rocha:2012jg, Peter:2012jh, Zavala:2012us, Vogelsberger:2014pda, Elbert:2014bma, Kaplinghat:2015aga, Fry:2015rta} are favored. On the other hand, the Bullet Cluster constrains self-interactions to be $\sigma_\text{scatter}/m_\chi<1.25\, \text{cm}^2/\text{g}$~\cite{Clowe:2003tk, Markevitch:2003at, Randall:2007ph}.

%%%%%%%%%%%%%%%%%%%%%%%%%%%FREEZE-IN%%%%%%%%%%%%%%%%%%%%%%%%%%%%%%%%%%%%%
%%%%%%%%%%%%%%%%%%%%%%%%%%%%%%%%%%%%%%%%%%%%%%%%%%%%%%%%%%%%%%%%%%%%%%%%
%%%%%%%%%%%%%%%%%%%%%%%%%%%%%%%%%%%%%%%%%%%%%%%%%%%%%%%%%%%%%%%%%%%%%%%
\subsection{Freeze-in Production}
If the dark matter is very weakly coupled to the thermal bath, e.g., a feebly interacting massive particle, or FIMP, it will never reach thermal equilibrium with the thermal bath. Assuming that the initial abundance of dark matter is negligible, then the feeble interaction with the bath leads to some dark matter production during the evolution of the universe. The abundance of the dark matter eventually freezes in with a yield that increases by increasing the interaction strength~\cite{Hall:2009bx}.
Dark matter can be produced via the freeze-in mechanism through different general scenarios such as decay ($B_1\rightarrow B_2\chi$, $B_1\rightarrow\chi\chi$), scattering ($B_1B_2\rightarrow B_3\chi$), and pair production ($B_1B_2\rightarrow\chi\chi$), where $B_i$'s are bath particles. The freeze-in mechanism could be relevant for dark matter particles of any mass smaller than the reheating temperature of the universe. Although most of the studies in the context of the freeze-in production mechanism focus on dark matter with a weak scale mass~\cite{Hall:2009bx, McDonald:2001vt, Choi:2005vq, Kusenko:2006rh, Petraki:2007gq}, superheavy dark matter production is also viable in a minimal scenario where the dark matter particles have only gravitational interactions with the thermal bath~\cite{Garny:2015sjg, Garny:2017kha}.

Considering pair production to be the dominant freeze-in production channel, then the number density of dark matter particles evolves according to the Boltzmann equation,
\begin{equation}
\dot{n}_\chi+3Hn_\chi\approx\int d\Pi_{B_1}d\Pi_{B_2}d\Pi_{\chi_1}d\Pi_{\chi_2}(2\pi)^4\delta^4(p_{B_1}+p_{B_2}-p_{\chi_1}-p_{\chi_2})|M_{B_1B_2\rightarrow\chi_1\chi_2}|^2 f_{B_1}f_{B_2}.
\end{equation}
By assuming that the masses of $B_1$ and $B_2$ are negligible compared to the dark matter mass, and assuming a constant matrix element $|M_{B_1B_2\rightarrow\chi_1\chi_2}|^2=\lambda^2$ for the interaction, the abundance is evaluated as~\cite{DEramo:2017ecx}, 
\begin{equation}
\Omega_\chi(\lambda) h^2\simeq\frac{405 \sqrt{5}\lambda^2}{16384\,\pi^{13/2}g_*^{3/2}}\frac{M_\text{Pl}s_0}{\rho_c}h^2.
\end{equation}

A population of PBHs can also give rise to feebly coupled dark matter particles via Hawking evaporation. Because of the negligible interactions of these particles, the products of the freeze-in process and the Hawking evaporation together form the final abundance of dark matter. Therefore
\begin{eqnarray}
\Omega_\chi^{\text{fr-in}}(\lambda)h^2+\Omega_\chi^{\text{BH}}(m_\chi,M_\text{BH},\beta)h^2\leq\Omega_\text{c}h^2.
\label{eq:FIMPY}
\end{eqnarray}

The PBH contribution to the relic abundance of FIMP dark matter requires the feeble coupling of these particles to the thermal bath to be even feebler, which makes searches for these particles even more challenging (see, eg.~\cite{Bernal:2017kxu}).

Note that Eq.~(\ref{eq:FIMPY}) is valid as long as evaporation happens in a radiation-dominated era. If the abundance of PBHs is large enough to initiate an early-matter dominated epoch, then there would be no $\beta$ dependence in Eq.~(\ref{eq:FIMPY}). In this case, PBH evaporation, except for a specific dark matter mass, underproduces dark matter and therefore it cannot be the only source of  dark matter.

%%%%%%%%%%%%%%%%%%%%%%%%%%%WIMPZILLA%%%%%%%%%%%%%%%%%%%%%%%%%%%%%%%%%%%%%
%%%%%%%%%%%%%%%%%%%%%%%%%%%%%%%%%%%%%%%%%%%%%%%%%%%%%%%%%%%%%%%%%%%%%%%%
%%%%%%%%%%%%%%%%%%%%%%%%%%%%%%%%%%%%%%%%%%%%%%%%%%%%%%%%%%%%%%%%%%%%%%%
\subsection{Gravitational Production: WIMPZILLA}
\label{subsec:zilla}
One intriguing possibility to go beyond the unitarity limit on the mass of thermally produced dark matter is that dark matter might be composed of non-thermal supermassive particles, e.g., so-called WIMPZILLAs~\cite{Chung:1998ua, Chung:1998zb, Chung:1998bt, Chung:2001cb, Chung:2004nh}. There are a variety of possible mechanisms for generating WIMPZILLAs in the early universe. 
One possibility is the standard gravitational particle creation by the expansion of the universe acting on quantum fluctuations of the vacuum~\cite{Chung:1998ua, Chung:1998zb, Chung:1998bt, Chung:2001cb, Chung:2004nh, Kuzmin:1998uv, Kuzmin:1998kk, Kuzmin:1999zk}. 
The novelty of this mechanism is its capability of producing dark matter particles with mass of the order of the inflaton mass even when the dark matter does not interact at all with other particles, not even the inflaton. The abundance of gravitationally produced WIMPZILLAs depends on the ratio $m_\chi/H_I$~\cite{Chung:1998zb} where $H_I\lesssim 2.5\times 10^{-5}M_\text{Pl}\sim 10^{14}\,\text{GeV}$ is the Hubble rate during  inflation~\cite{Akrami:2018odb}. The density  of WIMPZILLAs today can be comparable to the critical density of the universe for $0.04\lesssim m_\chi/H_I\lesssim 2$~\cite{Chung:1998zb}.

WIMPZILLAs can also be produced at the end of  inflation, during the reheating process~\cite{Chung:1998rq}. By solving a set of coupled Boltzmann equations for the inflaton field energy density, the radiation energy density, and the dark matter energy density, one can show that~\cite{Chung:1998rq} $\Omega_\chi\sim m_\chi^2 \langle\sigma v\rangle(2000\,T_\text{RH}/m_\chi)^7$, where $T_\text{RH}$ is the reheat temperature and $\langle\sigma v\rangle$ is the thermally-averaged annihilation cross section of the dark matter particles. In this scenario dark matter particles of mass as large as $\sim 10^3$ times the reheat temperature may be produced with the right abundance today.
 
Another possibility for producing particles with larger masses is the preheating process during which inflaton field oscillations at the end of inflation, in the regime of a broad parametric resonance, lead to the rapid creation of massive particles~\cite{Kofman:1994rk, Kofman:1997yn, Shtanov:1994ce}. In this scenario, the upper limits on the mass of bosonic and fermionic particles are $10^{16}\,\text{GeV}$~\cite{Chung:1998ua, Chung:1998zb, Chung:1998bt, Chung:2001cb, Chung:2004nh} and $M_\text{Pl}$~\cite{Giudice:1999fb, Chung:1999ve}, respectively. The production of WIMPZILLAs through preheating and their abundance are model dependent. For example, in the context of slow-roll inflation with potential $V(\phi)=m_\phi^2\phi^2/2$ with the inflaton coupled to the WIMPZILLA by a term $g \chi^2\phi^2/2$, the dark matter abundance depends on $m_\chi/H_I$ and $g M_\text{Pl}/H_I$~\cite{Chung:1998bt}. For a fixed value of $g M_\text{Pl}/H_I$, $\Omega_\chi$ is a decreasing function of $m_\chi/H_I$. Surprisingly, for a fixed value of  $m_\chi/H_I$, $\Omega_\chi$ is not a monotonic function of $g M_\text{Pl}/H_I$~\cite{Chung:1998bt}.

Finally, WIMPZILLAs may also be produced in bubble collisions in a first-order phase transition that completes inflation~\cite{La:1989za}. In this picture, bubble nucleation leads to the transition of the universe from a false vacuum state to the true vacuum state~\cite{Guth:1980zm}. The bubble walls, which contain huge potential energy originating from the false vacuum, expand and turn their potential energy into kinetic energy. The highly relativistic bubble walls eventually collide, and during the collision they can produce non-thermal dark matter particles with masses up to $\gamma m_\phi$, where $\gamma$ is the relativistic Lorentz factor~\cite{Kolb:1996jr, Kolb:1997mz}. If WIMPZILLAs are fermions that interact with the inflaton field through a Yukawa interaction of strength $g_{\chi\chi\phi}$, then the number of dark matter particles created during the bubble collisions is $N_\chi\sim f_\chi M_\text{Pl}/m_\chi$ where $f_\chi\simeq g^2\text{ln}(\gamma m_\phi/2m_\chi)$~\cite{Watkins:1991zt, Masiero:1992bv}.

PBHs can also produce WIMPZILLAs via Hawking evaporation. Due to the negligible non-gravitational interactions of these particles, the WIMPZILLAs produced by a cosmological mechanism and those produced by Hawking evaporation of PBHs together form the final abundance of dark matter, i.e.
\begin{eqnarray}
\Omega_\chi^{\text{WIMPZILLA}}(m_\chi, g)h^2+\Omega_\chi^{\text{BH}}(m_\chi,M_\text{BH},\beta)h^2\leq\Omega_\text{c}h^2.
\label{eq:WZY}
\end{eqnarray}

If there is WIMPZILLA production from PBH evaporation, fewer WIMPZILLAs must originate from other gravitational production mechanisms in order to avoid overclosing the universe. This, in turn, can affect current constraints on superheavy dark matter particles from direct detection searches~\cite{Albuquerque:2003ei} and indirect detection searches, such as the constraints on  decaying WIMPZILLAs from ultra high energy cosmic rays ~\cite{Alcantara:2019sco}, as well as isocurvature constraints~\cite{Chung:2004nh}.

Similar to the freeze-in case, Eq.~(\ref{eq:WZY}) is valid as long as PBHs evaporate during a radiation-dominated era. If the abundance of PBHs is large enough to initiate an early matter-dominated epoch, then there would be no $\beta$ dependence in Eq.~(\ref{eq:WZY}). In this case, PBH evaporation alone will underproduce dark matter, and therefore it cannot be the only source of the dark matter.

%%%%%%%%%%%%%%%%%%%%%%%%%%%%Parameter Space%%%%%%%%%%%%%%%%%%%%%%%%%%%%%%%%%%%%
%%%%%%%%%%%%%%%%%%%%%%%%%%%%%%%%%%%%%%%%%%%%%%%%%%%%%%%%%%%%%%%%%%%%%%%%
%%%%%%%%%%%%%%%%%%%%%%%%%%%%%%%%%%%%%%%%%%%%%%%%%%%%%%%%%%%%%%%%%%%%%%%%%
\section{Cosmological Constraints}
\label{sec:ps}
In this section we discuss the constraints on the parameters of interest in this study (the PBH mass $M_\text{BH}$, the dark matter particle mass $m_\chi$, and the initial abundance of PBHs $\beta$) from the CMB, BBN, structure formation, and capture of superheavy dark matter by PBHs. We use these constraints to find the available parameter space for dark matter production via PBH evaporation, which will be analyzed in section~\ref{sec:results}.
%%%%%%%%%%%%%%%%%%%%%%%%%%%%PBH Mass%%%%%%%%%%%%%%%%%%%%%%%%%%%%%%%%%%%%
%%%%%%%%%%%%%%%%%%%%%%%%%%%%%%%%%%%%%%%%%%%%%%%%%%%%%%%%%%%%%%%%%%%%%%%%
\subsection{PBH Mass}
The range of allowed PBH masses  is constrained from below by CMB observations and from above by BBN limits. The lightest viable black holes can form when the Hubble rate is less than or equal to the Hubble rate during inflation, $H_I\leq2.5\times 10^{-5}M_\text{Pl}$. 
Eq.~(\ref{eq:initialmass}) translates $H_I$ into a lower limit on the PBH mass,
\begin{equation}
M_\text{BH}>\frac{\gamma}{2}\frac{1}{2.5\times 10^{-5}}M_\text{Pl}\simeq\left(\frac{\gamma}{0.2}\right)0.1 \text{g}.
\end{equation}

To avoid spoiling the agreement between theoretical predictions and observational constraints, we demand that PBHs evaporate before BBN, $T_\text{eva}<T_\text{BBN}$. This leads to an upper limit on the black hole mass of $M_\text{BH}<10^9 \text{g}$.

We note also that when the PBHs cause an early matter-dominated era, 
so long as the PBH masses are $M_\text{BH}\lesssim 10^9\,\text{g}$, evaporation will lead to a radiation-dominated universe with $T_\text{RH-BH}>T_\text{BBN}$ in accordance with the standard cosmological model~\cite{Baumann:2007yr}.

%%%%%%%%%%%%%%%%%%%%%%%%%%%%Warm Dark Matter%%%%%%%%%%%%%%%%%%%%%%%%%%%%%%%%%%%%
%%%%%%%%%%%%%%%%%%%%%%%%%%%%%%%%%%%%%%%%%%%%%%%%%%%%%%%%%%%%%%%%%%%%%%%%
\subsection{Warm Dark Matter}
Another constraint on dark matter produced from PBH evaporation comes from the requirement that the dark matter should be cold enough to avoid erasing small-scale structures via free-streaming. In the absence of dark matter interactions with the thermal bath, the only way that dark matter particles can lose energy is by red-shifting. Following~\cite{Fujita:2014hha}, and knowing that the average energy of the produced particles is of the order of the initial temperature of the black hole, the redshifted momentum of a dark matter particle today is related to its initial momentum by 
\begin{equation}
p_0=\frac{a_\text{eva}}{a_0}p_\text{eva}\sim\frac{a_\text{eva}}{a_0}\bar E.
\end{equation}
Assuming $a_0=1$, we can connect $a_0$ to $a_\text{eva}$ by using the scale factor at matter-radiation equality $a_\text{eq}=\Omega_r/\Omega_m$:
\begin{equation}
p_0=\frac{a_\text{eva}}{a_\text{eq}}\frac{\Omega_r}{\Omega_m}\bar E=\left(\frac{\rho_r(T_\text{eq})}{\rho_r(T_\text{eva})}\right)^{1/4}\frac{\Omega_r}{\Omega_m}\bar E.
\end{equation}
Then using Eq.~(\ref{eq:Teva}) and $\rho_r(T_\text{eq})=\rho_c/a_\text{eq}^3$, the expression for $p_0$ can be recast as
\begin{equation}
\left(\frac{p_0}{1\,\text{GeV}}\right)\simeq2.5\times10^{-12}\left(\frac{M_\text{BH}}{M_\text{Pl}}\right)^{1/2}.
\end{equation}
An upper bound on the typical velocity of warm dark matter today can be obtained from the lower bound on the mass of a thermal warm dark matter particle~\cite{Fujita:2014hha}. 
Taking the warm dark matter to be heavier than
$3.5\, \text{keV}$~\cite{Irsic:2017ixq}, one finds $v_\chi \lesssim1.8 \times 10^{-8}$~\cite{Masina:2020xhk} as the upper bound on the velocity, so
\begin{equation}
\left(\frac{m_\chi}{1\,\text{GeV}}\right)\gtrsim 1.6\times 10^{-4}\left(\frac{m_\text{PBH}}{M_\text{Pl}}\right)^{1/2}.
\label{eq:hot}
\end{equation}
%%%%%%%%%%%%%%%%%%%%%%%%%%%%Capture of Superheavy Dark Matter by PBHs%%%%%%%%%%%%%%%%%%%%%%%%%%%%%%%%%%%%
%%%%%%%%%%%%%%%%%%%%%%%%%%%%%%%%%%%%%%%%%%%%%%%%%%%%%%%%%%%%%%%%%%%%%%%%
\subsection{Capture of Superheavy Dark Matter by PBHs}
\label{sec:capt}
Production of superheavy dark matter through freeze-in or gravitational production is only efficient at the end of inflation and the onset of reheating. PBHs formed in the early universe might capture  preexisting superheavy dark matter particles before the PBHs evaporate. If PBHs are abundant enough to initiate an early matter-dominated epoch $(\beta>\beta_c)$, PBH evaporation will underproduce dark matter and another production mechanism would be necessary. But a large initial abundance of PBHs increases the dark matter capture rate and worsens the underproduction problem. To see when capture becomes important, we compare the total number of dark matter particles $N_\chi$ within a Hubble volume $V_H$ to the number of  dark matter particles $N_{\chi-\text{Capt}}$ captured during a Hubble time $t_H$~\cite{Stojkovic:2004hz},  
\begin{equation}
\frac{N_{\chi-\text{Capt}}}{N_\chi}=\frac{(n_\chi\sigma_{\chi,\text{BH}} v_\chi t_H) n_\text{BH} V_H}{n_\chi V_H}=\frac{n_\text{BH}\sigma_{\chi,\text{BH}} v_\chi}{H}.
\label{eq:captrate}
\end{equation}
Here $n_\chi$ is the number density of dark matter particles, $n_\text{BH}$ is the number density of PBHs, $v_\chi$ is the velocity of a dark matter particle, $H$ is the Hubble rate, and $\sigma_{\chi,\text{BH}}$ is the cross section for gravitational capture of a non-relativistic massive particle by a black hole of Schwarzschild radius $r_\text{S}$, given by~\cite{Frolov:1998wf}
\begin{equation}
\sigma_{\chi,\text{BH}}=\frac{4\pi r_\text{S}^2}{v_\chi^2}.
\label{eq:crosssec}
\end{equation}
Using Eqs.~(\ref{eq:captrate}), (\ref{eq:crosssec}), (\ref{eq:beta}), and (\ref{eq:initialmass}) we have
\begin{equation}
\frac{N_{\chi-\text{Capt}}}{N_\chi}=\frac{3\gamma}{v_\chi}\beta\simeq6\times10^3\left(\frac{\gamma}{0.2}\right)\left(\frac{10^{-4}}{v_\chi}\right)\beta,
\end{equation}
where $v_\chi \sim 10^{-4}$ is a typical value for the average velocity of a superheavy dark matter particle, taken from the benchmark spectrum provided by~\cite{Chung:1999ve} for a fermionic dark matter particle of mass $m_\chi\sim 10^{18}\,\text{GeV}$ produced gravitationally. Therefore, for a typical abundance $\beta\sim10^{-4}$, the capture of already-existing superheavy dark matter particles by PBHs is significant, $N_{\chi-\text{Capt}}/N_\chi\sim1$.   
%%%%%%%%%%%%%%%%%%%%%%%%%%%%Results%%%%%%%%%%%%%%%%%%%%%%%%%%%%%%%%%%%%
%%%%%%%%%%%%%%%%%%%%%%%%%%%%%%%%%%%%%%%%%%%%%%%%%%%%%%%%%%%%%%%%%%%%%%%%
\section{Results}
\label{sec:results}
\begin{figure}[ht]
  \centering
  \includegraphics[width=0.95\textwidth]{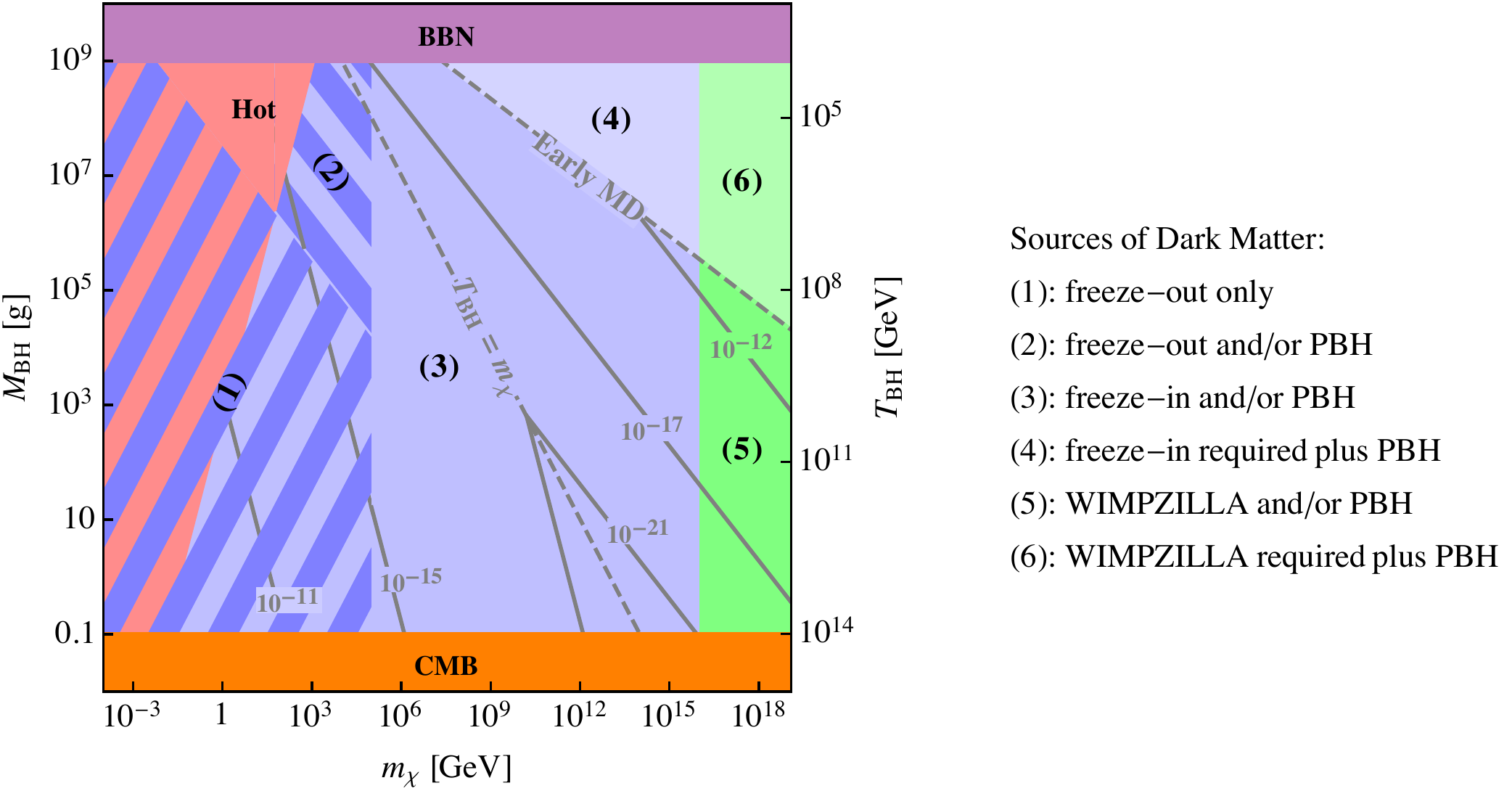}
  \caption{Constraints on Majorana fermion dark matter production by PBH evaporation in the $(m_\chi,M_{\rm BH})$ plane. Solid grey lines show the upper limits on $\beta$ assuming PBHs produce all of the dark matter. The orange, purple, and salmon shaded regions are excluded by constraints from the CMB, BBN, and structure formation (``Hot'' dark matter), respectively.
Above the early matter-dominated line ('Early MD'), $\beta$ is so large that it leads to an early matter-dominated epoch, and the abundance of dark matter particles produced by PBH evaporation is independent of $\beta$. The dark blue stripes depict the regions of parameter space where the freeze-out production mechanism is also possible; in region (1) PBHs evaporate before freeze-out and in region (2) evaporation happens after freeze-out. In the light blue region, the freeze-in production mechanism is possible; in region (3) evaporation happens in a radiation-dominated era while in region (4) it occurs in an early matter-dominated era. In the green region, gravitational production mechanisms other than PBH evaporation are possible; in region (5) PBHs evaporate in a radiation-dominated epoch while in region (6) evaporation happens in an early matter-dominated epoch. The legend to the right of the figure shows the interplay between dark matter production by PBH evaporation and other sources of dark matter in each region.}
  \label{fig:fig1}
\end{figure}

\begin{figure}[t]
  \centering
  \includegraphics[width=0.45\textwidth]{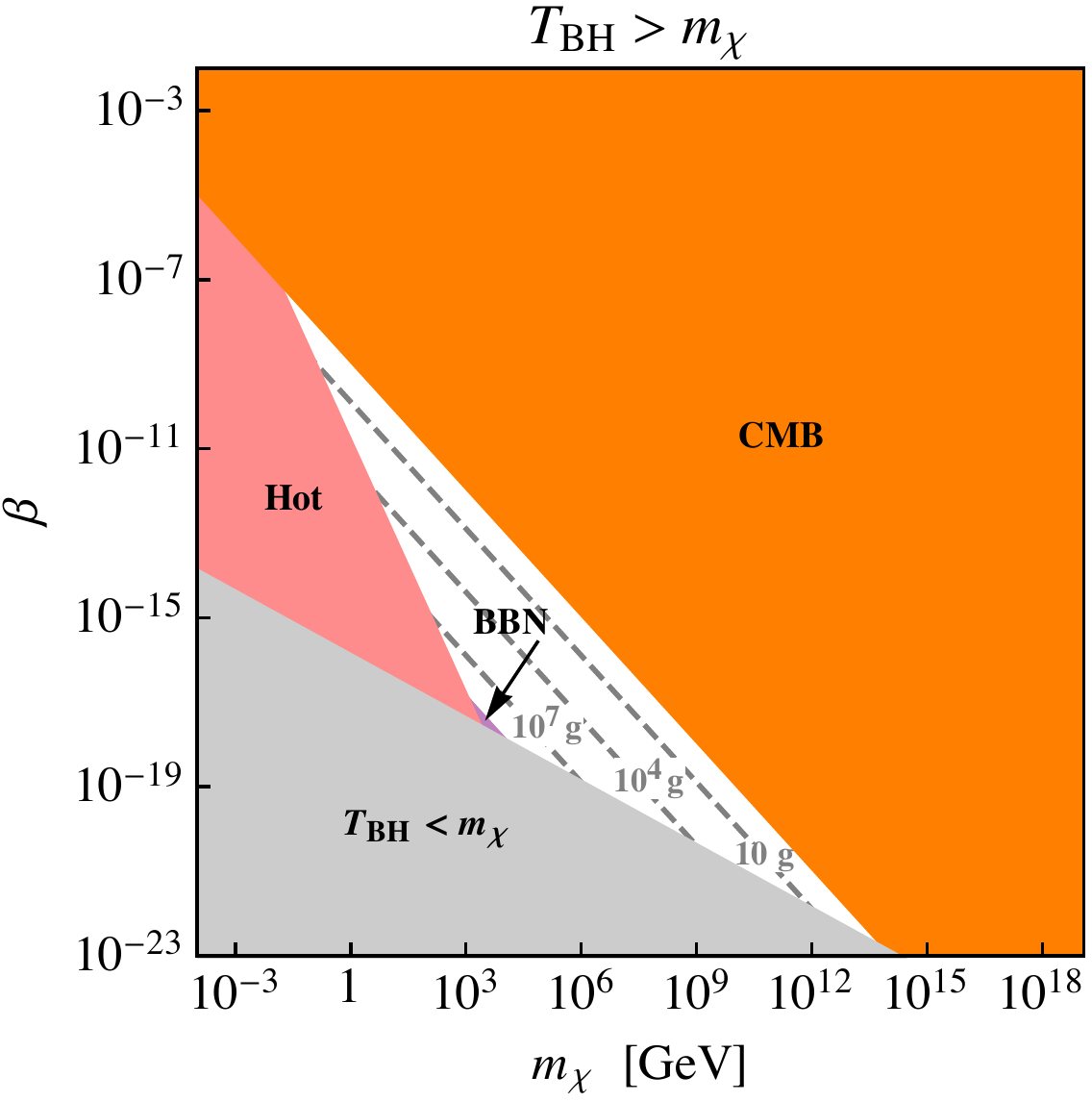}
    \includegraphics[width=0.45\textwidth]{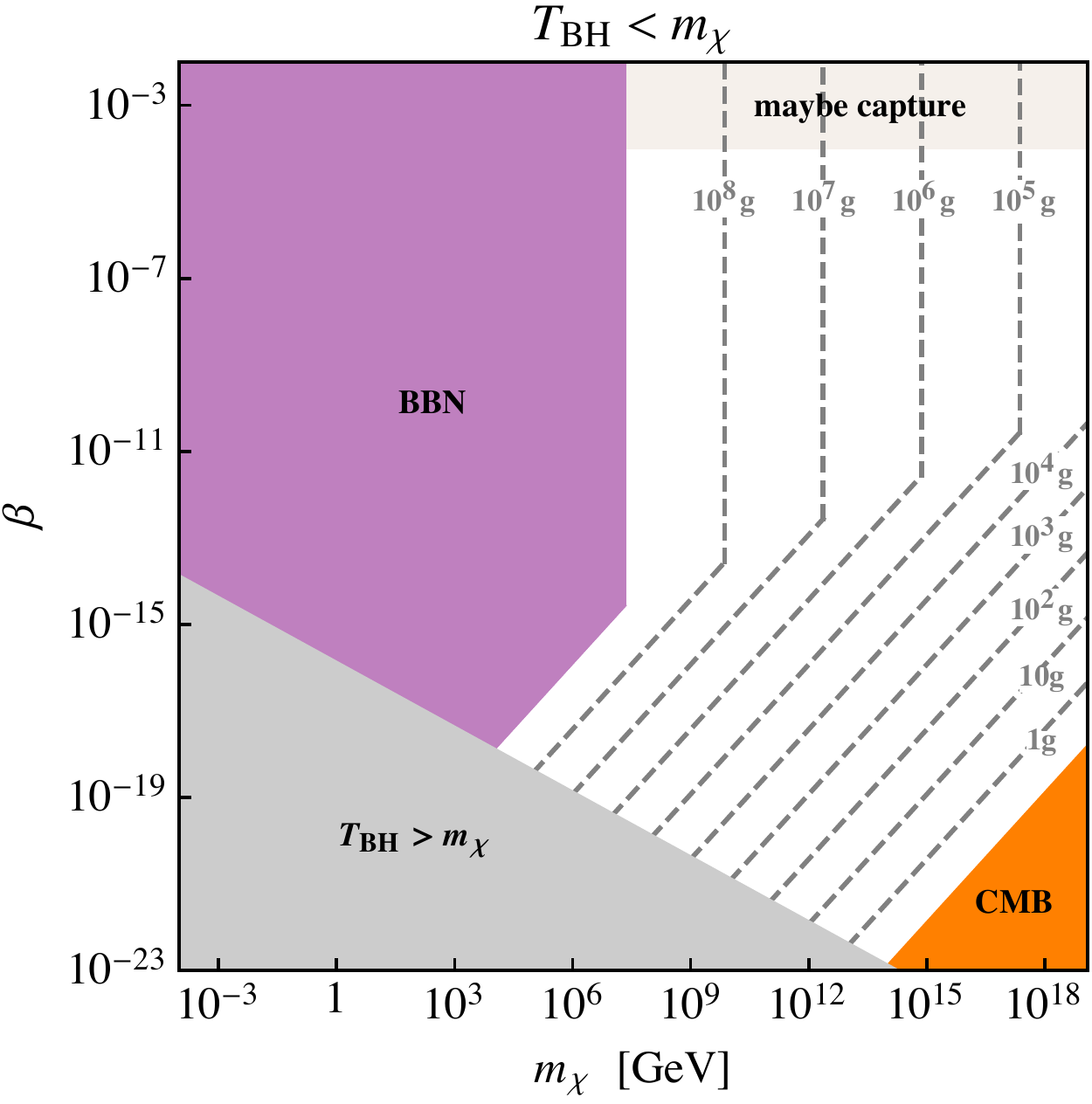}
  \caption{Constraints on dark matter production by PBH evaporation in the $(m_\chi,\beta)$ plane when $T_\text{BH}>m_\chi$ (left panel) and $T_\text{BH}<m_\chi$ (right panel). Gray dashed lines in left (right) panel show the upper (lower) limits on $M_\text{BH}$ assuming PBHs produce all of the dark matter.}
  \label{fig:fig2}
\end{figure}

Our primary results are collected in Figs.~\ref{fig:fig1} and \ref{fig:fig2}, which are described in detail below. Fig.~\ref{fig:fig1} shows the upper limits on the initial abundance of PBHs, $\beta$, as well as other possible mechanisms for obtaining the relic abundance of dark matter, and regions of the $(m_\chi, M_\text{BH})$ parameter space excluded by various cosmological constraints as described in Section~\ref{sec:ps}.

%%%%%%%%%%%%%%% FIGURE %%%%%%%%%%%%%%%%%%%%%
%%%%%%%%%%%%%%% FIGURE %%%%%%%%%%%%%%%%%%%%%

If all the dark matter in the universe is explained by dark matter produced in PBH evaporation, then one can express the upper limit on the initial abundance $\beta$ of PBHs as a function of their mass $M_\text{BH}$ and the mass $m_\chi$ of the dark matter particles by equating the dark matter abundance produced by PBH evaporation $\Omega_\chi^{\text{BH}}$ to the observed dark matter abundance $\Omega_c$,
\begin{equation}
\Omega_\chi^{\text{BH}}\left[m_\chi,M_\text{BH},\beta(m_\chi,M_\text{BH})\right]=\Omega_c.
\end{equation}
The solid grey contours in Fig.~\ref{fig:fig1} represent the upper limits on $\beta$ in the $(m_\chi, M_\text{BH})$ plane when PBHs produce all the dark matter (assuming Majorana fermion dark matter) while evaporating in a radiation-dominated universe. When the required $\beta$ is so large $(\beta>\beta_c)$ that it leads to an early matter-dominated epoch (above the ``Early MD'' line in the upper right corner), then the dark matter abundance is independent of $\beta$, cf. discussion at the end of subsection~\ref{subsec:DMfromPBH}.
The excluded regions by CMB, BBN, and structure formation (``Hot'') are shaded in orange, purple, and salmon, respectively. In all other regions of Fig.~\ref{fig:fig1}, it is possible to obtain the observed abundance of dark matter.

We overlay in Fig.~\ref{fig:fig1} the regions favored by other mechanisms of dark matter production, such as freeze-out ($10^{-4}\, \text{GeV}\lesssim m_\chi\lesssim 10^5\, \text{GeV}$), freeze-in ($m_\chi\lesssim 10^{16}\, \text{GeV}$) and gravitational production of WIMPZILLAs ($10^{16}\, \text{GeV}\lesssim m_\chi\lesssim M_\text{Pl}$). These mechanisms may/must (or may not) contribute to the total abundance of dark matter as described in the plot legend.  Before continuing, we discuss the features of each region.

In the freeze-out domain, when PBH evaporation happens before the freeze-out of the dark matter, region (1), all effects of PBHs, such as dark matter production, a possible transition to an early matter-dominated era, etc., will be erased by thermalization. Therefore, in region (1), $\beta$ can obtain any value less than one; the freeze-out mechanism alone sets the final abundance of dark matter.
Due to thermalization, relativistic dark matter particles produced by PBHs can exchange momentum with other particles and cool down. As a result, even very small $m_\chi$ becomes possible, noted by the diagonal blue stripes that extend into the excluded ``Hot'' region.
 
In region (2), PBH evaporation happens after dark matter freeze-out.
In this case, the value of $\beta$ must be less than its upper limit (solid grey contours), and the annihilation cross section of dark matter must be at least as large as the generic annihilation cross section expected for thermal dark matter alone.
In region (2) the abundance of the dark matter today can be explained in three ways: by the freeze-out mechanism only, by PBH evaporation only, or by a combination of the two.

Region (3) is the freeze-in domain, when PBHs do not lead to an early matter-dominated epoch. Here
PBHs and the freeze-in mechanism both contribute directly to the final abundance of dark matter. Hence the value of $\beta$ must be less than its upper limit, and the coupling between dark matter and the thermal bath must be correspondingly smaller than the typical value in the general freeze-in scenario. In region (3) the abundance of the dark matter today can be explained by freeze-in only, by PBHs only, or by a combination of the two.

Region (4) is also in the freeze-in domain, but here the large abundance of
PBHs causes an early matter-dominated epoch.
In region (4), the dark matter from PBH evaporation is underproduced. In this region, dark matter can be explained by the freeze-in mechanism only or by a combination of PBH evaporation and the freeze-in mechanism together. Though the amount of dark matter produced by PBH evaporation in region (4) is independent of $\beta$, to avoid capturing dark matter by PBHs the value of $\beta$ cannot be arbitrarily large, cf.~Subsection.~\ref{sec:capt}.

Regions (5) and (6) are the WIMPZILLA domain, without and with an early matter-dominated epoch, respectively.
In region (5), PBHs and the gravitational production mechanism of WIMPZILLAs both contribute directly to the final abundance of dark matter. Hence the value of $\beta$ must be less than its upper limit, and the controlling parameters of gravitational production of WIMPZILLAs should be adjusted accordingly. In region (5) the abundance of the dark matter today can be explained by gravitational production mechanism only, by PBHs only, or by a combination of the two.

In region (6), the dark matter from PBH evaporation alone is underproduced, and, similar to region (4), its amount is independent of $\beta$. Therefore in region (6) dark matter can be explained by the gravitational production mechanism only or by a combination of PBHs and the gravitational production mechanism. Again, to avoid excessive capture of
dark matter by PBHs, the value of $\beta$ cannot be arbitrarily large, cf.~Subsection.~\ref{sec:capt}.

In Fig.~\ref{fig:fig2} we focus on dark matter production by PBHs only. In the left (right) panel, which corresponds to $T_\text{BH}>m_\chi$ ($T_\text{BH}<m_\chi$), we display the value of the upper (lower) limit on $M_\text{BH}$ that results in $\Omega_c h^2$ today. Depending on the mass of the PBH, $\beta$ can vary from $10^{-23}$ to $10^{-7}$ ($10^{-23}$ to $10^{-4}$) when $T_\text{BH}>m_\chi$ ($T_\text{BH}<m_\chi$). For the case $T_\text{BH}<m_\chi$, a large $\beta$ may increase the probability of capturing recently-produced superheavy dark matter particles from PBH evaporation by PBHs that are still around. As such, a large initial abundance of PBHs ($\beta\gtrsim10^{-4}$) should be treated carefully.

%%%%%%%%%%%%%%% FIGURE %%%%%%%%%%%%%%%%%%%%%
%%%%%%%%%%%%%%% FIGURE %%%%%%%%%%%%%%%%%%%%%

%%%%%%%%%%%%%%%%%%%%CONCLUSIONS AND OUTLOOK%%%%%%%%%%%%%%%%%%%%%%%%%
%%%%%%%%%%%%%%%%%%%%%%%%%%%%%%%%%%%%%%%%%%%%%%%%%%%%%%%%%%%%%%%%%
%%%%%%%%%%%%%%%%%%%%%%%%%%%%%%%%%%%%%%%%%%%%%%%%%%%%%%%%%%%%%%%%%
\section{CONCLUSIONS AND OUTLOOK}
\label{sec:conclusion}
In this paper we have explored the effects of PBH evaporation, as a novel dark matter production mechanism, on well-motivated particle dark matter scenarios, including freeze-out, freeze-in, and gravitational production of dark matter. Production of dark matter particles via Hawking evaporation of PBHs accompanied by other dark matter production mechanisms (gravitational or non-gravitational) can lead to interesting model-independent modifications in these mechanisms. We have shown that these modifications alter the prospects for dark matter detection, which we summarize below.

We have demonstrated that for the freeze-out mechanism, thermalization erases the effects of PBHs when they evaporate before the freeze-out of the dark matter, and consequently the freeze-out mechanism controls the final abundance of the dark matter. On the other hand, when PBHs evaporate after freeze-out, they contribute directly into the final abundance of dark matter. A contribution to the dark matter abundance from PBH evaporation diminishes the required contribution from the freeze-out mechanism, which is feasible if the annihilation cross section of dark matter particles increases.  In fact, the PBH contribution can become so large that the annihilation cross section saturates the unitarity bound. A larger annihilation cross section strengthens the signals in indirect detection searches for dark matter.

For mechanisms lacking thermalization due to a feeble coupling to the thermal bath, such as freeze-in or gravitational production of superheavy dark matter, we have shown that when the presence of PBHs  does not lead to an early matter-dominated epoch, the contribution to the dark matter abundance from PBH evaporation decreases the required contribution of other production mechanisms. This manifests itself in an even feebler coupling that makes the direct detection searches, indirect detection searches, and collider searches for these particles even more challenging. We have also found that if PBHs do cause an early matter-dominated epoch, the dark matter from PBH evaporation is underproduced, and its amount is independent of the initial abundance of PBHs. Therefore, in this scenario, evaporation of PBHs alone cannot explain the whole abundance of dark matter today, and an additional production mechanism, e.g.~freeze-in or gravitational production of dark matter, is required.
 
PBHs are extremely fascinating targets for theoretical and phenomenological studies. Irrespective of the underlying model of dark matter, the contribution of Hawking evaporation of PBHs to the abundance of dark matter today is inevitable. Understanding the interplay between PBHs and the alternative sources of dark matter can help us to learn more about the early universe, the nature of the dark matter and its detection prospects.

%%%%%%%%%%%%%%%%%%%%%%%%%%%%%acknowledgments%%%%%%%%%%%%%%%%%%%%%%%%%%
%%%%%%%%%%%%%%%%%%%%%%%%%%%%%%%%%%%%%%%%%%%%%%%%%%%%%%%%%%%%%%%%%%
%%%%%%%%%%%%%%%%%%%%%%%%%%%%%%%%%%%%%%%%%%%%%%%%%%%%%%%%%%%%%%%%%%
\acknowledgments
This work has been supported in part by NSF grant PHY-1720282.
%%%%%%%%%%%%%%%%%%%%%%%%%%%%%%%%%%%%

\end{document}